\documentclass[10pt,conference]{IEEEtran}
\IEEEoverridecommandlockouts
\usepackage{cite}
\usepackage{makecell}
\usepackage{amsmath,amssymb,amsfonts}
\usepackage{algorithmic}
\usepackage{graphicx}
\usepackage{textcomp}
\usepackage{xcolor}
\usepackage{url}
\usepackage{listings}
\usepackage{todonotes}
\usepackage{xcolor}
\usepackage{hyperref}
\usepackage{multirow}
\usepackage{tcolorbox}
\usepackage{arydshln}
\usepackage[commandnameprefix=always,final]{changes}
\usepackage{graphicx}
\usepackage{caption}
\usepackage{subcaption}
\usepackage{marginnote}
\usepackage{orcidlink}

\definecolor{codegreen}{rgb}{0,0.6,0}
\definecolor{codegray}{rgb}{0.5,0.5,0.5}
\definecolor{codepurple}{rgb}{0.58,0,0.82}
\definecolor{backcolour}{rgb}{0.95,0.95,0.92}

\newcommand{\pipeline}{\textbf{Pipeline}}
\newcommand{\nn}{\textbf{NN}}

\newcommand{\synthetic}{\emph{Synthetic 1}}

\newcommand{\synthetictwo}{\emph{Synthetic 2}}

\newcommand{\real}{\emph{Real}}

\newcommand{\annotated}{\emph{Annotated}}

\newcommand{\fbeta}{F-$\beta$}

\newcommand{\realpy}{\emph{Real-Py150}}

\newcommand{\realpypi}{\emph{Real-PyPI}}

\newtcolorbox{tbox}[3][]
{
    colframe = #2!25,
    colback  = #2!10,
    coltitle = #2!20!black,  
    title    = {#3},
    #1,
}

\lstdefinestyle{mystyle}{
    backgroundcolor=\color{backcolour},   
    commentstyle=\color{codegreen},
    keywordstyle=\color{blue},
    escapechar={|},
    numberstyle=\tiny\color{codegray},
    stringstyle=\color{codepurple},
    basicstyle=\ttfamily\footnotesize,
    breakatwhitespace=false,         
    breaklines=true,                 
    captionpos=b,                    
    keepspaces=true,
    numbersep=5pt,                  
    showspaces=false,                
    showstringspaces=false,
    showtabs=false,                  
    tabsize=2
}

\begin{document}

\title{The Power of Types: Exploring the Impact of \\ Type Checking on Neural Bug Detection \\ in Dynamically Typed Languages

\thanks{Partially supported by the FRQNT-B2X project (file number: 319955), IT30340 Mitacs Accelerate, and the Wallenberg AI, Autonomous Systems
and Software Program (WASP), Sweden}
}

\author{\IEEEauthorblockN{Boqi Chen \orcidlink{0000-0002-1451-3603}}

\IEEEauthorblockA{\textit{ECE} \\
\textit{McGill University}\\
Montreal, Canada}
\and
\IEEEauthorblockN{Jos\'e Antonio Hern\'andez L\'opez \orcidlink{0000-0003-2439-2136}}
\IEEEauthorblockA{\textit{IDA} \\
\textit{Link\"oping University}\\
Link\"oping, Sweden}
\and
\IEEEauthorblockN{Gunter Mussbacher \orcidlink{0009-0006-8070-9184}}
\IEEEauthorblockA{\textit{ECE} \\
\textit{McGill University}\\
Montreal, Canada}
\and
\IEEEauthorblockN{D\'aniel Varr\'o \orcidlink{0000-0002-8790-252X}}
\IEEEauthorblockA{
\textit{IDA / ECE} \\
\textit{Link\"oping University} / 
\textit{McGill University}\\
Link\"oping, Sweden / Montreal, Canada}
}

\maketitle

\begin{abstract}
\emph{[Motivation]} Automated bug detection in dynamically typed languages such as Python is essential for maintaining code quality. The lack of mandatory type annotations in such languages can lead to errors that are challenging to identify early with traditional static analysis tools. Recent progress in deep neural networks 
has led to increased use of neural bug detectors. In statically typed languages, a type checker is integrated into the compiler and thus taken into consideration when the neural bug detector is designed for these languages.


\emph{[Problem]} However, prior studies overlook this aspect during the training and testing of neural bug detectors for dynamically typed languages. When an optional type checker is used, assessing existing neural bug detectors on bugs easily detectable by type checkers may impact their performance estimation. 
Moreover, including these bugs in the training set of neural bug detectors can shift their detection focus toward the wrong type of bugs.




\emph{[Contribution]}
We explore the impact of type checking on various neural bug detectors for variable misuse bugs, a common type targeted by neural bug detectors. Existing synthetic and real-world datasets are type-checked to evaluate the prevalence of type-related bugs. Then, we investigate how type-related bugs influence the training and testing of the neural bug detectors.


\emph{[Findings]}
Our findings indicate that existing bug detection datasets contain a significant proportion of type-related bugs. Building on this insight, we discover integrating the neural bug detector with a type checker can be beneficial, especially when the code is annotated with types.
Further investigation reveals neural bug detectors perform better on type-related bugs than other bugs.
Moreover, removing type-related bugs from the training data helps improve neural bug detectors' ability to identify bugs beyond the scope of type checkers.




\end{abstract}

\begin{IEEEkeywords}
type checking, neural bug detection, dynamically typed languages
\end{IEEEkeywords}

\section{Introduction}

Bug detection is a crucial task in software engineering (SE). Studies suggest that between 0.5 to 25 bugs may occur per 1000 lines of industrial code~\cite{mcconnell2004code,habib2018many}. The implications of these bugs in software systems can be profound, leading to outcomes ranging from minor system malfunctions to potential loss of human lives~\cite{fonseca2017empirical,jemal2022presence,poulsen2004software,zhivich2009real}.
Bug detection is even more important in dynamically typed languages such as Python and Javascript, which are more error-prone than statically typed languages~\cite{ray2014large}. Furthermore, the absence of mandatory type checking in these languages often leads to more type-related bugs compared to statically typed languages~\cite{gao2017type,khan2021empirical}. 

Automated bug detection aims to identify bugs early in the software development cycle. Traditionally, this task has been depending on static analysis which interprets the semantics of the source code \cite{aftandilian2012building,calcagno2015moving,hovemeyer2004finding}. Recent advances in machine learning (ML) have increased interest in using neural networks for identifying various bugs by learning implicit bug patterns from faulty programs \cite{habib2019neural}. These bug detectors typically aim to identify bugs, such as variable misuse~\cite{allamanis2017learning} or binary operator bugs~\cite{troshin2022probing}, in \emph{syntactically correct} programs. Consequently, their training and evaluation datasets comprise both correct and faulty programs that are syntactically valid. 





\begin{figure}[tb]
    \centering
    \includegraphics[width=0.8\linewidth]{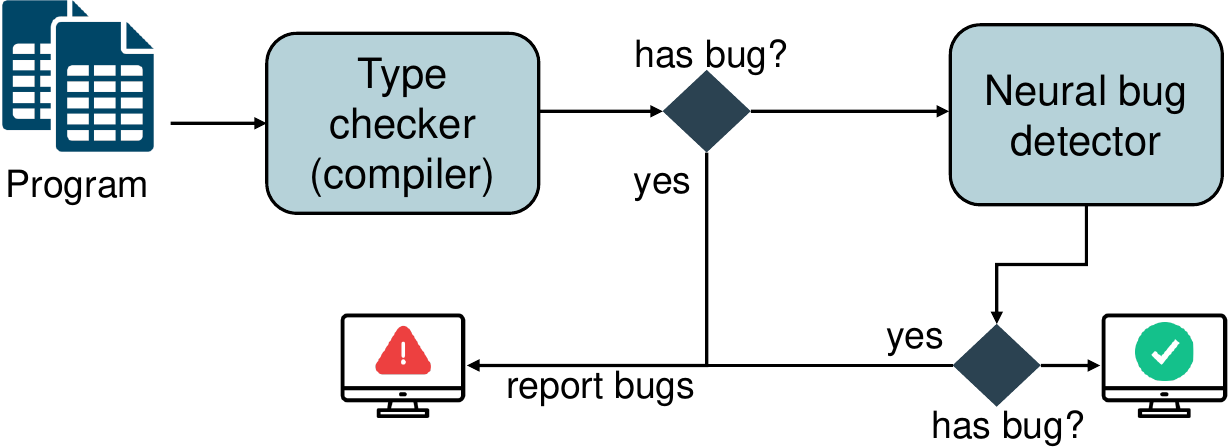}
    \caption{Use of an NBD for statically typed languages}
    \label{fig:motivation}
\end{figure}

In statically typed languages, such as Java and C\#, type checking is integrated into the compiler's program analysis process. Hence, previous research on neural bug detectors (NBDs) for these languages has considered this aspect \cite{allamanis2017learning}. \autoref{fig:motivation} shows how developers can use a type checker together with an NBD for a statically typed language. The process begins with type checking; if a bug is identified by the type checker (categorized as \emph{type-related bug} in our paper), 
it is immediately reported and rectified. When the type checker does not find any bugs, the NBD evaluates the program. It reports any discovered bugs, otherwise the program is deemed correct. Hence, \emph{neural bug detectors focus on programs for which the type checker does not detect any bugs}.


Thus, the training and evaluation of NBDs must be contextualized to ensure their relevance and efficiency.
First, NBD models should be evaluated exclusively on bugs not detectable by the type checker to ascertain their true effectiveness.
Moreover, the training of NBDs should prioritize real-world performance to identify bugs that elude type checkers, thus enhancing their usefulness.




However, when studying NBDs for dynamically typed languages, such as Python, previous research often disregards this crucial aspect due to the absence of inherent static type checkers~\cite{VasicKMBS19,kanade2020learning,hellendoorn2019global}. As a result, little is known about the impact of type-checking on the performance of these NBDs.
Still, the growing popularity of optional type annotations in these languages, which facilitate static type checking, suggests a shift towards integrating type checking in the development workflow \cite{khan2021empirical,di2022evolution}. Moreover, compared to NBDs, traditional type checkers (1) generate more explainable outputs, (2) provide soundness guarantees with well-studied theoretical frameworks, and (3) require less computational resources and no training data. When a type checker is used, for example, as part of continuous integration, existing performance assessments may not accurately represent the NBD's effectiveness in such settings. Besides, training NBDs on such bugs may inadvertently skew the model's detection capability towards less relevant types. 



This paper explores the impact of type checking on 
contemporary NBDs in the context of Python, a widely used dynamically typed language. Our analysis encompasses a broad spectrum of neural architectures, including Gated Graph Neural Networks (GGNN) \cite{li2015gated}, GREAT \cite{hellendoorn2019global}, CodeBERT~\cite{feng2020codebert}, GraphCodeBERT~\cite{guo2020graphcodebert}, and UniXcoder~\cite{guo2022unixcoder}. 

Our study investigates variable misuse bugs, which are frequently targeted by NBDs as they are widespread in SE practice \cite{rabin2021understanding,richter2023train,tarlow2020learning}. For instance, 12\% of the bugs in the ManySStuBs4J corpus are variable misuse bugs~\cite{karampatsis2020often}, while around 6\% of build errors in Google engineering systems can be attributed to this type of bugs~\cite{allamanis2022graph}. Additionally, Allamanis et al.~\cite{allamanis2017learning} find such bugs in RavenDB, a popular NoSQL database, highlighting their prevalence in open-source projects.

We assess how type-related bugs affect the training and evaluation of NBDs using both synthetic and real-world datasets. 
Specifically, we address the following research questions:

\noindent \textbf{RQ1:} \emph{How prevalent are type-related bugs in variable misuse datasets?}

\noindent \underline{Motivation}:
Before examining how type checking influences the training and evaluation of NBDs, it is imperative to assess if bugs identifiable by a type checker exist within popular variable misuse datasets.


\noindent \underline{Results}: We find that a notable proportion of bugs can be identified by a type checker in synthetic and real-world datasets that include \emph{programs that are not annotated with types}. This proportion is even higher when programs are type annotated.


\noindent \textbf{RQ2:} \emph{How does type checking influence the performance of NBDs when they are used together?}

\noindent \underline{Motivation}: 
Since existing variable misuse datasets contain a notable proportion of type-related bugs identifiable by a type checker, the aim of RQ2 is to evaluate how effective it is to use NBDs together with a type checker in terms of performance.

\noindent \underline{Results}: 
We find that, when the programs are not annotated with types, incorporating a type checker enhances the performance of GGNN and GREAT. Using a type checker with some other neural networks results in increased recall (more bugs detected) but decreased precision (more false alarms). However, when incorporating a type checker on annotated programs, both recall and precision consistently improve.



\noindent \textbf{RQ3:} \emph{How do type-related bugs influence the performance evaluation of NBDs?}

\noindent \underline{Motivation}: 
When incorporating a type checker into the development process, NBDs should prioritize detecting bugs that elude type checkers. Hence, understanding their effectiveness in identifying this category of bugs is crucial.

\noindent \underline{Results}: We observe that NBDs exhibit notably better performance on type-related bugs. 



\noindent \textbf{RQ4:} \emph{How do type-related bugs influence the training of NBDs?}

\noindent \underline{Motivation}: Since NBDs tend to perform better on type-related bugs, the existence of such bugs in the training data may bias the NBD towards the wrong type of bugs already caught by the type checker. Therefore, in RQ4, we investigate how filtering out type-related bugs impacts the training of NBDs.

\noindent \underline{Results}: We find that NBDs trained on programs excluding type-related bugs perform better in identifying bugs beyond the scope of type checkers on synthetic and real-world datasets. We also observe a decrease in the precision of bug detection. An improvement in joint performance is observed when the recall of bugs is prioritized.

\noindent\textbf{Added value.} Our findings offer practical insights into the optimal use of a type checker with an NBD for dynamically typed languages, particularly 
when recall for variable misuse bugs is more important than precision, or when type annotations are present. Integrating a type checker with an NBD necessitates excluding type-related bugs from evaluation to mitigate the potential of overly optimistic results. Likewise, during training, it is advisable to exclude these bugs to improve bug detection. 

\noindent\textbf{Artifacts.} Artifacts for this study are available online~\cite{artifacts_icse25}.

\noindent \textbf{Organization.}
This paper is organized as follows: 
\autoref{sec:background} introduces the background and context of our study. Following this, \autoref{sec:existance} examines the prevalence of type-related bugs in popular datasets for variable misuse. \autoref{sec:effect} evaluates the impact of type checking on the training and evaluation of NBDs. Subsequently, \autoref{sec:discusssion} explores the potential implications of our findings. Finally, \autoref{sec:rel-work} reviews related literature, and \autoref{sec:conclusion} concludes the paper. 




\section{Background}
\label{sec:background}
\subsection{Type checking for Python}
Type checkers are a category of bug detectors based on static analysis targeting to identify defects using the types of variables. These type checkers are typically integrated into the compiler of statically typed languages and can be used as external tools in dynamically typed languages \cite{rak2020python}. 

Python is a dynamically typed language thus it only determines the type of a variable at runtime. However, starting with Python 3.5, optional type annotations are introduced to both help the developer understand the code better and enable static type checkers to detect type-related bugs \cite{PEP484}. 

There are two type-checking frameworks in Python: gradual typing and type inference \cite{rak2020python}. Many popular type checkers, such as \emph{mypy} \cite{mypy}, operate on the principles of \emph{gradual typing}, a framework that detects inconsistencies in types using explicit type annotations in the program \cite{siek2007gradual}. However, gradual typing often assigns \emph{unknown} types to unannotated variables, which limits its ability to detect type-related bugs in the absence of type annotations. \emph{Type inference} is a technique similar to gradual typing but with a key distinction: it infers types for variables even without explicit type annotation in the code, thereby enabling the detection of a broader spectrum of type-related bugs. 
Given the differences between the two frameworks, this paper uses \texttt{pytype} \cite{pytype}, a popular open-source type inference-based type checker, when the program is not annotated with types, and uses both \texttt{pytype} and \texttt{mypy} when type annotations exist.

Recent advances have seen deep learning-based approaches being applied to type inference in Python. The core concept involves using neural networks to learn typing patterns implicitly in programs, subsequently annotating variables based on these patterns. Tools such as DeepTyper \cite{hellendoorn2018deep} and Typilus \cite{allamanis2020typilus} have demonstrated the ability to infer complex types that are difficult to infer by traditional methods. However, these tools have not yet been widely adopted in practice due to their high computational demands and the challenges in explaining the rationale behind their type inferences. Thus, we do not consider these types of inference tools in this paper.

\subsection{Variable misuse bugs}

A variable misuse bug occurs when a developer unintentionally uses an incorrect variable. \autoref{lst:codeExp} shows an example of a variable misuse bug. In this snippet, the developer mistakenly uses the variable \texttt{first} instead of \texttt{last} in line 8, \texttt{if (assn[1]!=first[1])}. Unlike in statically typed languages, this program will not cause a \textit{syntactic error} but will raise a \textit{runtime error} when the function is executed.

\begin{lstlisting}[language=python, caption={Example synthetic faulty program extracted from a Python variable misuse dataset. Incorrect variable use location is bold and colored in red}, label={lst:codeExp}, numbers=left]
    def take_last_assignment(source):
        |\textbf{first}|=True
        last=None
        for assn in source:
            if |\textbf{first}|:
                last=assn
                |\textbf{first}|=False
            if (assn[1]!=|\textcolor{red}{\textbf{first}}|[1]):
                (yield last)
            last=assn
        if (last is not None):
            (yield last)
\end{lstlisting}


In this paper, we exclusively focus on variable misuse bugs due to their prevalence in SE practice~\cite{rabin2021understanding,richter2023train,tarlow2020learning}, and being commonly targeted by NBDs~\cite{VasicKMBS19,allamanis2017learning, hellendoorn2019global,troshin2022probing}. Furthermore, some of these bugs can potentially be caught by type checkers, which makes this kind of bugs a good candidate for investigating NBDs in a practical SE setting. We focus on detecting a single bug in a code snippet, which is a common assumption in the literature of neural bug detection \cite{VasicKMBS19, hellendoorn2019global,troshin2022probing}.

\subsection{NBDs for variable misuse}


This section overviews the task of detecting variable misuse, and the neural network architectures used to tackle such task.

\subsubsection{Task}
Given a program, a bug detector targeting variable misuse bugs produces two outputs: (1) a decision on whether the program contains such a bug and (2) the location of the variable misuse, if the program contains a bug. Formally, given a bug detector $\mathfrak{D}$ and a program (function) $P$, the output of the bug detector can be represented as $\mathfrak{D}(P) = (d, l)$, where $d$ is a Boolean value indicating whether the program has a bug or not, and $l$ represents the location of the bug when $d = \texttt{true}$.

\subsubsection{Architectures}
The three prevalent neural network architectures used to tackle the variable misuse task are transformers, graph neural networks (GNN), and hybrid approaches.

\noindent\textbf{Transformers} represent a class of neural networks that process inputs as sequences of tokens, leveraging a multi-head attention mechanism to capture global relationships within these tokens \cite{vaswani2017attention}. 
In this case, the programs are represented as a sequence of tokens and the output bug location ($l$) corresponds to the token that causes the bug. While it is possible to train transformers from scratch with randomly initialized weights, they are typically pre-trained on extensive data corpora and subsequently fine-tuned for specific tasks. These pre-trained models are known as pre-trained language models. This paper focuses on encoder-only language models as they are predominantly used in fine-tuning for classification problems~\cite{lin2022survey}. Specifically, in our experiments, we use CodeBERT \cite{feng2020codebert}, GraphCodeBERT \cite{guo2020graphcodebert}, and UniXcoder~\cite{guo2022unixcoder}, all of which are pre-trained models specialized in source code.

\noindent\textbf{Graph neural networks (GNNs)} process graphs with both structural and feature information \cite{kipf2016semi}. 
Hence, the programs are depicted as program graphs~\cite{allamanis2017learning}, encompassing details concerning control flows (\textit{e.g.,} branching and loops) and data flows (\textit{e.g.,} last read and last write of a variable). Consequently, the output bug location ($l$) corresponds to the node representing the token responsible for the bug. The most frequently used GNNs for bug detection are gated graph neural networks (GGNN)~\cite{allamanis2017learning} which use a gated recurrent unit (GRU) \cite{cho2014properties} to perform the neighbor aggregation~\cite{li2015gated}. 

\noindent\textbf{Hybrid approaches}
combine both transformer and graph neural network approaches. In our study, we explore the Graph-Relational Embedding Attention Transformer (GREAT) architecture proposed by Hellendoorn et al.~\cite{hellendoorn2019global}. GREAT harnesses the capabilities of both transformers and GNNs by enhancing the attention mechanism with biases for different relationship types. The central concept behind GREAT is incorporating edge information into the transformer architecture by using a \textit{relation type bias} term in the attention computation.

\section{Existence of Type-Related Bugs}
\label{sec:existance}

This section summarizes why type checking is important to be used for detecting Python bugs, and examines the presence of type-related bugs in popular synthetic and real-world datasets. Additionally, it discusses the methodology for assessing the prevalence of type-related bugs in these datasets.


\subsection{Motivation}

Previous work in creating NBDs for Python has largely overlooked the potential presence of type-related bugs in their datasets due to the lack of type annotations in Python code. However, recent type checkers such as \texttt{pytype} \cite{pytype} based on \textit{type inference} can identify type-related bugs even in the absence of explicit type annotations. For example, in \autoref{lst:codeExp}, the variable usage \texttt{first} at line 8 can be detected by a type checker through type inference. In this context, \texttt{first} is expected to be a Boolean variable. Attempting member access for a boolean variable is invalid, resulting in an \textit{unsupported operand error} for the program.

In order to understand the influence of such defects on the performance of NBDs, the first step is to assess the existence of such bugs in datasets used for training and evaluation. We evaluate existing synthetic and real-world bug detection datasets to address the following research question:

\begin{itemize}
   \item \textbf{RQ1:} \emph{How prevalent are type-related bugs in variable misuse datasets?}
\end{itemize}

\subsection{Evaluation setup}
In this section, we present our methodology for conducting type checks on datasets designed for Python variable misuse bug detection. We detail the datasets used in our study and provide an overview of our approach.

\subsubsection{Datasets}
We analyze two widely used synthetic bug detection datasets and two real-world datasets. Furthermore, we manually annotate a subset of the real-world datasets.

\textbf{ETH Py-150 synthetic bugs (\synthetic{})} \cite{hellendoorn2019global} is a dataset specifically crafted for the detection and correction of variable misuse bugs within Python functions. This dataset is synthetically generated from the ETH Python-150 dataset \cite{kanade2020learning} by substituting correct variable usages with incorrect ones, chosen randomly within the same function. For each faulty function, the dataset specifies the bug's location, a list of candidate variables for repair, and the correct variable in the candidates. The dataset contains 1.7M training functions, 200k for validation, and 950k for evaluation. 

\textbf{Synthetic training set (\synthetictwo{})} \cite{he2022distribution} is a training dataset constructed following a similar approach as \synthetic{}. Instead of directly creating synthetic bugs from the programs in the ETH Py150 dataset, this dataset is generated by extracting programs directly from the open-source projects used to construct Py150. Specifically, repositories without a commit fix related to real-world variable misuse bugs are used. In total, around 147k samples were constructed for both faulty and correct functions from 2654 repositories.

\textbf{Py150 Real Bugs (\realpy{})} \cite{he2022distribution} is a dataset containing real variable misuse bugs extracted from open-source Python projects used to construct the Py150 dataset by applying a set of heuristics to identify commits fixing these bugs. However, the amount of faulty programs is insufficient to train a NBD. We include the correct programs from the test splits as well as faulty programs from all splits. 
In total, the dataset contains 1,292 bugs and 61,539 correct programs



\textbf{PyPI Real Bugs (\realpypi{})} \cite{allamanis2021self} contains 2436 faulty functions with different types of bugs found in all 285k packages from the Python Package Index (PyPI). We only filter out the variable misuse bugs and use them for evaluation. This dataset provides commits to the fix of the bug. We use this information to extract original faulty programs from project repositories. While some projects used in the dataset were deleted, we were able to extract 1,051 samples of variable misuse bugs from this dataset. We merge this dataset with the \realpy{} dataset to create the merged \textbf{\real{}} dataset.

\textbf{Annotated dataset (\annotated{})}
While it is suggested to integrate type annotation and checking into the development workflow \cite{khan2021empirical,di2022evolution}, most programs in the datasets are still not annotated with types. Previous work has shown that type checkers may not identify all errors in unannotated programs \cite{xu2023well}. To better understand the effect of type annotation in this case, we manually annotate a subset of the \real{} dataset, referred to as \annotated{}. Aligned with previous work \cite{khan2021empirical,gao2017type}, we randomly select 400 programs from the \real{} dataset sourced from Github for manual annotation. We choose 200 correct programs and 200 buggy programs to balance the annotated dataset. Next, we present the annotation process.

\subsection{Type annotation process}
\begin{figure}[tb]
    \centering
    \includegraphics[width=0.8\linewidth]{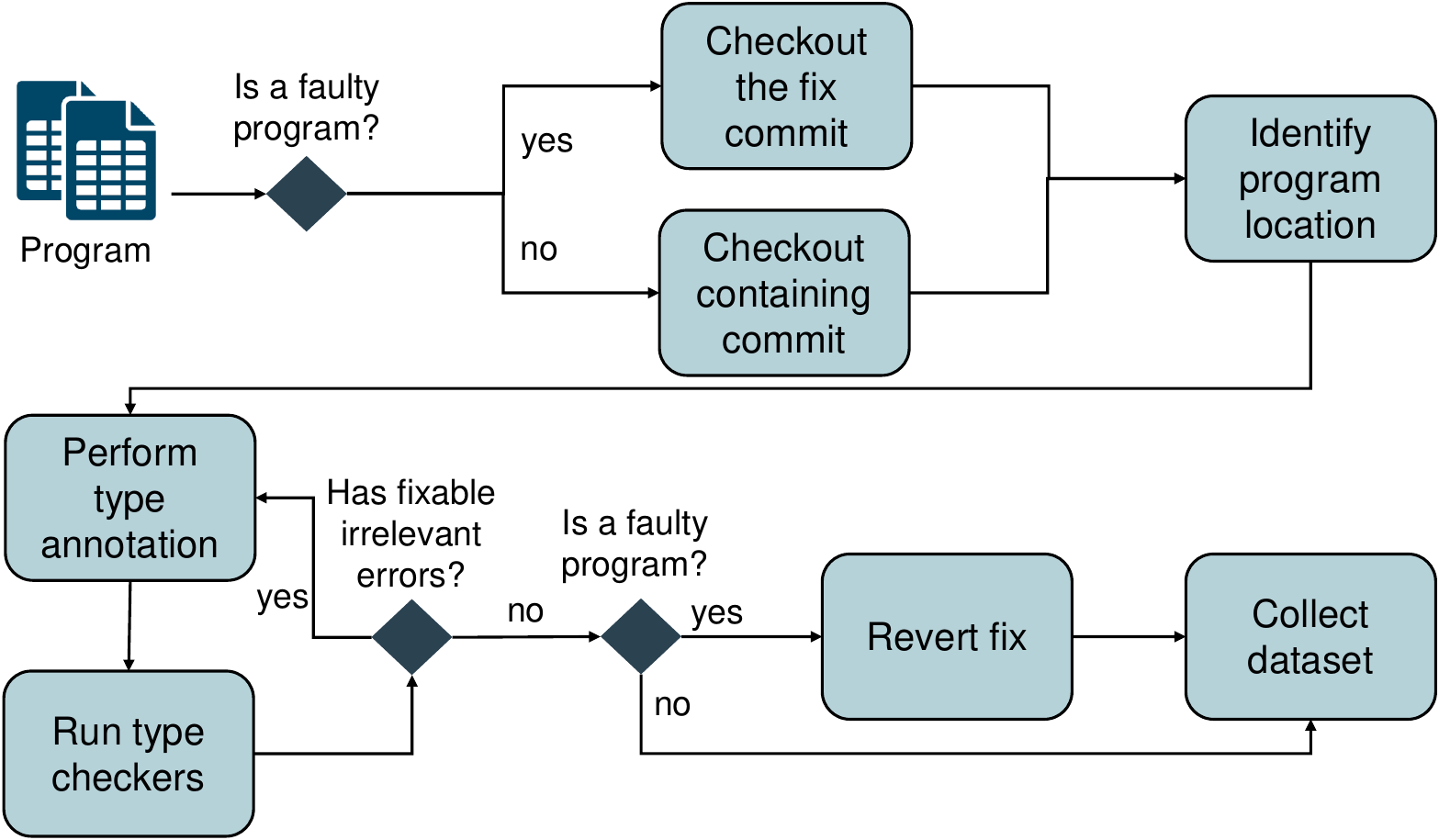}
    \caption{Type annotation process for the \real{} dataset}
    \label{fig:annotation-process}
\end{figure}

We adopt an annotation process similar to previous work on type checking for Python \cite{khan2021empirical}. \autoref{fig:annotation-process} demonstrates this process for which the \emph{entire repository} is used. To ensure consistency, one author of the paper annotated all programs, and the results were discussed among all authors to resolve any discrepancies. For a faulty program, the version from the \emph{fix commit} is used. For a correct program, we identify the most recent commit containing the \emph{exact same program} as of the dataset's publication date. This step ensures that the program used during the annotation is correct, preventing bias in the procedure due to the existence of bugs in the program.

Once the correct version is set, the file of the function to be annotated is located using the metadata. Then, type annotations are added to \emph{all} variables and functions used. If necessary, class and function stubs are added to the file to provide type annotations for symbols used but not defined in the function.  

After the initial type annotation, both \texttt{mypy} and \texttt{pytype} are run to check for any irrelevant errors. If such errors can be fixed by refining the type annotations, the annotations are updated accordingly. This process is repeated until no errors are present or the remaining errors cannot be fixed by refining annotations, such as false alarms from the type checker.

Finally, the fix commit is reverted for faulty programs so that any further errors identified by the type checker can be considered type-related bugs. Both the function and the defined stubs are collected for evaluation.


\subsection{Addressing the RQ}
\begin{figure}
    \centering
    \includegraphics[width=0.65\linewidth]{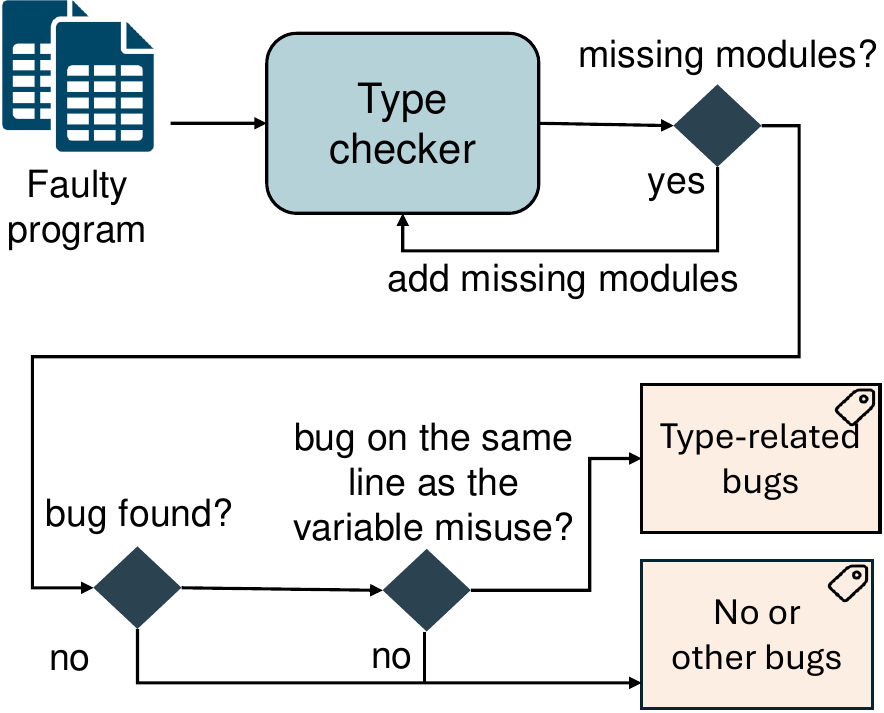}
    \caption{An overview of the type-related bugs labeling process}
    \label{fig:bug-label}
\end{figure}

For each faulty code snippet in the datasets, we use 
a type checker to detect type-related bugs within the function. The unannotated code is only type-checked with \texttt{pytype} while the annotated programs are checked both with \texttt{pytype} and \texttt{mypy}. Preprocessing and filtering are performed to ensure the bugs caught by the type checker are indeed the variable misuse bugs of interest. \autoref{fig:bug-label} shows an overview of the process.



\textbf{Preprocessing:}
Given that each unannotated code snippet within the datasets only contains one function and omits the imported packages for that function, the type checker may detect undesired \textit{import errors}. To mitigate such issues, our approach includes a dual-phase of type checking. Initially, the first phase of type checking is conducted to identify any absent packages. Subsequently, all detected missing packages are imported at the beginning of the code snippet. This step is not necessary for the annotated programs since the missing symbols are already added as stubs.

\textbf{Type checking and filtering:}
Then a second phase of type checking is executed. To account for potential irrelevant issues raised by the type checker, we ignore all errors related to imported packages and internal errors from \texttt{pytype} such as \textit{pyi errors} indicating that \texttt{pytype} cannot produce type inference for the program. Finally, only errors detected on the same line as the variable misuse bug location are considered identifiable by the type checker. 

We conduct type checking on 
all datasets, quantifying the proportion of bugs identifiable by a type checker. We refer to these bugs identifiable by a type checker as \textit{type-related bugs}.

\subsection{Results}
     
     
     




\begin{figure*}[tb]
    \centering
    \includegraphics[width=0.8\linewidth]{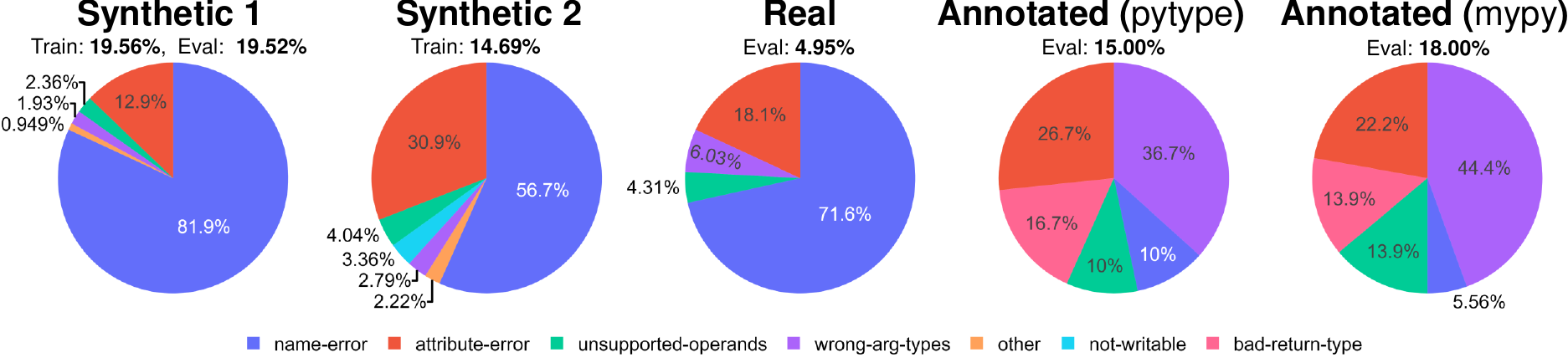}
    \caption{Distribution of type-related bugs in the synthetic, real-world, and annotated datasets. In \synthetic{}, we plot only the distribution of the evaluation split (the training split follows a similar distribution).}
    \label{fig:error-type}
\end{figure*}

The top of \autoref{fig:error-type} shows the percentage of bugs detectable by a type checker 
for all four datasets. For the \annotated{} dataset, both results from \texttt{pytype} and \texttt{mypy} are presented.

In general, all datasets contain a notable fraction of type-related bugs. Synthetic datasets contain the highest fraction, with roughly $19.56\%$ in \synthetic{} and $14.89\%$ in \synthetictwo{}. As these datasets are frequently used for training, this fraction could potentially lead NBDs to deviate their focus towards type-related bugs, which is undesirable in cases where developers use type checkers.

The type checker can also detect a notable portion of $4.95\%$ of real-world bugs in \real{}. When annotations are added to the programs, the estimated percentage of detected bugs increases significantly to around $15.00\%$ for \texttt{pytype} and $18.00\%$ for \texttt{mypy}. Our results are consistent with previous work \cite{khan2021empirical}, which indicates that \texttt{mypy} can identify around $15\%$ of bugs in annotated Python programs. Our slightly higher recall may be attributed to focusing on variable misuse bugs.


The bottom of \autoref{fig:error-type} also offers a more detailed analysis of type-related errors, illustrating the distribution of such bugs in both synthetic and real-world datasets. Since \texttt{mypy} gives different category names, we manually convert these names to the same as in \texttt{pytype}.  
In general, there are five types of bugs that appear most frequently in \textit{unannotated datasets}:

\begin{itemize}
    \item \textit{name-error} (56.7\%-81.9\%) often arises when a variable name is used without prior declaration or in parallel with the declaration. For example, the definition and use of a variable are in different branches of an if-else statement.
    \item \textit{attribute-error} (12.9\%-30.9\%) occurs when an attribute is accessed on a variable that does not have that attribute. For example, accessing an attribute of an integer variable.
    \item \textit{wrong-arg-types} (1.93\%-6.03\%) is raised when a function is called with arguments of wrong types. This is typically related to a built-in function while type annotation exists, such as using \texttt{len} to access the length of an integer.
    \item \textit{unsupported-operand} (2.36\%-4.31\%) is raised when an unsupported operation is performed on variables. For example, attempting to add a string to an integer.
    \item \textit{not-writable} (3.36\% in \synthetictwo{} only) is raised when trying to modify an immutable variable, such as an immutable tuple.
\end{itemize}


Most type-related bugs in unannotated programs are linked to trivial errors, \textit{e.g.,} undefined variables in \emph{name-errors}. Consequently, NBDs may excel in handling these instances, potentially inflating overall performance metrics. This poses a significant threat to evaluation, as these bugs may never be analyzed by the NBDs if a type checker is incorporated. 

The error distribution of \annotated{} differs from that of the unannotated datasets. Most detected bugs are \emph{attribute-errors} and \emph{wrong-arg-types}, which are closely related to variable types. Manual type annotation aids type checkers in detecting these errors more effectively.
Indeed, none of the new errors detected through added annotations are \emph{name-errors}. Also, a new error type, \emph{bad-return-type}, has been identified. This error is related to the function return type, thus requiring manual annotation in the function signature for detection.

\begin{tcolorbox}[left=1mm,right=1mm,top=1mm,bottom=1mm]
\textbf{RQ1}: A significant fraction of bugs 
are detectable by a type checker, ranging from $14.69\%$ to  $19.56\%$ for the synthetic datasets and $4.95\%$ for real-world bugs, even when programs are not annotated. This fraction becomes even higher for annotated real-world programs, ranging from $15.00\%$ to $18.00\%$ depending on the type checker.
Moreover, the training datasets contain a notable amount of detected bugs, potentially affecting the training of NBDs.

\end{tcolorbox}
\section{Effect of Type Checking}
\label{sec:effect}

\subsection{Motivation}

Neural bug detection datasets contain a significant proportion of type-related bugs, indicating the potential application of type checkers in detecting these bugs. In this section, we investigate how and when a type checker should be used together with an NBD for dynamically typed languages. Moreover, in scenarios where a type checker is used, NBDs are expected not to analyze bugs detectable by the type checker. Thus, to offer recommendations for deploying NBDs alongside a type checker, it is crucial to understand the impact of type-related bugs on the training and testing of these neural systems. Hence, we address the following three research questions:




\begin{itemize}
    \item \textbf{RQ2:} How does type checking influence the performance of neural bug detection when they are used together?
    \item \textbf{RQ3:} How do type-related bugs influence the evaluation of NBDs?
    \item \textbf{RQ4:} How do type-related bugs influence the training of NBDs?
\end{itemize}


\subsection{Experiment setup}
We introduce the neural networks used in the NBDs in our experiments. This is followed by datasets and evaluation metrics used to investigate the research questions.

\begin{table}[tb]
    \centering
    \caption{Bug Detection Neural Networks used in the experiments}
    \begin{tabular}{cccc}
    \hline
    \textsc{Model} & \textsc{Architecture} & \textsc{Training strategy} \\
    \hline
    GGNN \cite{allamanis2017learning} & GNN & From scratch \\

    CodeBERT \cite{feng2020codebert} & Transformer & Fine-tuning \\

    GraphCodeBERT \cite{guo2020graphcodebert} & Transformer & Fine-tuning \\

    UniXcoder \cite{guo2022unixcoder} & Transformer & Fine-tuning \\

    GREAT \cite{hellendoorn2019global} & Hybrid & From scratch \\
    \hline
    \end{tabular}
    \label{tab:models}
\end{table}

\subsubsection{NBDs}

The first column of \autoref{tab:models} shows the NBDs used in our experiments. The second column illustrates the architectures corresponding to each considered model. We cover three widely employed architectures: transformers, GNNs, and hybrid methods. The third column denotes if the detector is trained from scratch (\textit{i.e.,} initialized with random weights) or fine-tuned (\textit{i.e.,} initialized with a pre-trained neural network).



\subsubsection{Datasets}
We adopt the methodology from prior studies concerning training with a synthetic dataset and assessment using both the synthetic test split from the same dataset and real-world datasets~\cite{hellendoorn2019global,habib2019neural,he2022distribution,richter2023train}. Specifically, we select the \synthetic{} dataset as our training dataset because of its larger volume of data, necessary for training NBDs. The trained NBDs are then evaluated on (both annotated and unannotated) samples from the \real{} dataset.

Since the real-world datasets are constructed independently from the \synthetic{} dataset, some of the real-world bug programs may also appear in the training set of \synthetic{}. To avoid such data leakage issues, all potential duplicated samples are excluded from the real-world datasets during evaluation. Specifically, we use the metadata of each sample as a heuristic and remove all program samples that are from the same file in the same repository and have the same function signature as any samples from the synthetic training set. In total, 42.62\% correct programs and 7.89\% of bugs in the \realpy{} dataset, and 0.48\% of faulty programs in the \realpypi{} dataset are removed due to potential duplication with the training data. 


%
%

%
%




\subsubsection{Metrics}
In this work, we evaluate the performance of NBDs as a classification problem. Compared to standard classification metrics, we define stricter true positive cases for a bug detection task, taking into account bug localization. Specifically, given a set of faulty programs $\mathcal{P}_f$ and a bug detector $\mathfrak{D}$, the true positives ($TP$) are defined as the number of programs where the bug detector correctly identifies \emph{and} localizes the bug:
$$    TP(\mathfrak{D}, \mathcal{P}_f) = |\{p\in\mathcal{P}_f|d_{\mathfrak{D}(p)} = \texttt{true} \land l_{\mathfrak{D}(p)} = l_p\}|$$
where $d_{\mathfrak{D}(p)}$ represent decision and $l_{\mathfrak{D}(p)}$ represent the bug location of the program $p$ predicted by the NBD. Then, all other programs $p$ with $d_{\mathfrak{D}(p)} = \texttt{true}$ are considered as \emph{false positives} ($FP$). 
Moreover, false negatives ($FN$) are the number of faulty programs where the bug detector wrongly identifies the program as bug-free. Formally, the false negative ($FN$) is defined as the following:
$$    FN(\mathfrak{D}, \mathcal{P}_f) = |\{p\in\mathcal{P}_f|d_{\mathfrak{D}(p)} = \texttt{false}\}|$$
Finally, all other programs the NBD identifies as bug-free are considered as \emph{true negative} ($TN$). 

To evaluate the performance with NBDs, we use standard precision and recall as metrics:
$$    \text{Precision} = \frac{TP}{TP + FP}\:\:\:\:\text{Recall} = \frac{TP}{TP + FN}$$

To evaluate the joint performance of the bug detector, the F-score is typically used. However, the F-1 score treats precision and recall equally. In a real-world scenario, a programmer may prefer a higher recall to ensure all bugs are detected, even if it means more false positives, or a higher precision to reduce the number of false alarms. Therefore, we consider the \fbeta{} score~\cite{goutte2005probabilistic}, which allows us to adjust the importance of precision and recall. Specifically, the \fbeta{} score is defined as:
$$
    \text{F-}{\beta} = (1 + \beta^2) \times \frac{\text{Precision} \times \text{Recall}}{\beta^2 \times \text{Precision} + \text{Recall}}$$

while $\beta$ represents the relative importance of recall compared to precision. A higher $\beta$ signifies more emphasis on recall. When $\beta = 1$, the \fbeta{} score is equivalent to the F-1 score. In an SE context, a higher $\beta$ is preferable when detecting and fixing bugs are costly and time-consuming, while a lower $\beta$ is appropriate when developers need to spend much time on checking false alarms of automated detection tools.

To measure the differences in performance of the two settings, we use \emph{ratio change}. Given the performance values of two settings $p_a$ and $p_b$, the ratio change $\Delta$ of $p_b$ with respect to $p_a$ is calculated as follows:
\begin{equation*}
    \Delta = \frac{p_b - p_a}{p_a} \times 100
\end{equation*}

\noindent\textbf{Comparison to existing metrics.}
Existing work often uses correct program recall and localization accuracy as the metrics for an NBD \cite{hellendoorn2019global,habib2019neural} to evaluate bug classification and localization separately. In fact, when treating bug detection as a classification problem, localization accuracy can be seen as the recall in our metrics, while correct program recall is covered by the true negative rate. Furthermore, the \fbeta{} score takes both metrics into account and can provide a more comprehensive evaluation of the NBDs' end-to-end performance.

\subsubsection{Training settings}
Previous work often trains the NBD with the joint training objective of both bug detection and repair \cite{VasicKMBS19,allamanis2017learning,hellendoorn2019global}. In our preliminary experimentation, we notice that training without the repairing loss results in either similar or degraded performance compared to the joint objective. Consequently, we opt to use the joint training objective, as done in previous studies.

When training GGNN and GREAT models, we use the same hyperparameter configurations as in previous work~\cite{hellendoorn2019global} and train for 5 and 10 epochs, respectively. For fine-tuning, we train each model for 1 epoch and use a learning rate of $1e{-5}$ with a linear learning rate scheduler.

\subsection{RQ2: Influence of type checker}
\label{sec:influence-type-check}
\subsubsection{Addressing the RQ}
In this section, we aim to (A) evaluate the effect of integrating a type checker into the NBD similar to the ones used for statically typed languages as shown in \autoref{fig:motivation} and (B) investigate when and how the type checker should be used together with the NBD depending on the requirement of the developer. 

All NBDs are trained with the synthetic dataset on the variable misuse task. To gain insights on the influence of type checkers in a practical scenario, we evaluate the performance of the bug detectors on both the unannotated \emph{Real} and the annotated subset \annotated{}. In general, for each NBD, we consider two settings: (1) \textbf{NN}: the NBD is used alone and (2) \textbf{Pipeline}: the NBD is used together with a type checker in a pipeline as shown in \autoref{fig:motivation}. The type checker used in \pipeline{} is \texttt{pytype} for the \real{} dataset, as it can handle unannotated programs. For \annotated{}, we also evaluate \texttt{mypy} to understand the impact of different type checkers. Since the neural networks are trained on code without annotations, type annotations are removed when running the NBDs.

We also evaluate the precision and recall of the type checker using the \textit{Real} dataset. Specifically, we define $TP$ as the count of buggy samples correctly identified by the type checker (\textit{i.e.,} bugs identified in \autoref{fig:error-type}). $FN$ represents buggy samples missed by the type checker. The $FP$ is determined by running type checkers on the correct programs. A correct program is classified as $FP$ if the type checker raises a type of bug that was found by the type checker in faulty programs (e.g., \textit{name-error} or \textit{attribute-error} in \autoref{fig:error-type}).
This criterion is adopted because the type checker's capability extends to detecting various bug types beyond variable misuse.


\subsubsection{Results for the \textit{Real} dataset}

\begin{figure*}[ht]
  \centering
  \begin{subfigure}[b]{0.45\linewidth}
    \centering
    \footnotesize{
    \begin{tabular}{|c|cc|cc|}
    \hline
    \multirow{2}{*}{\textbf{Model}} & \multicolumn{2}{c|}{Precision \%} & \multicolumn{2}{c|}{Recall \%} \\
    \cline{2-5}
    & \textbf{NN} & \textbf{Pipeline} & \textbf{NN} & \textbf{Pipeline}\\
    \hline
    CodeBERT      & \textbf{27.91} & 24.54 & 32.29 & \textbf{34.83}\\
    GraphCodeBERT & \textbf{30.80} & 26.48 & 32.71 & \textbf{35.02}\\
    UniXcoder     & \textbf{31.71} & 27.38 & 34.54 & \textbf{36.96}\\
    GGNN          & 10.51 & \textbf{11.02} & 17.47 & \textbf{21.44}\\
    GREAT         &  9.86 & \textbf{10.05} & 26.59 & \textbf{29.83}\\
    \hline
    \texttt{pytype} & \multicolumn{2}{c|}{17.83} & \multicolumn{2}{c|}{5.01} \\
    \hline
    \end{tabular}
    }
    \caption{Precision and recall of the bug detectors on the unannotated \real{} dataset. Numbers are in percentage. The pipeline setting always improves recall but may reduce precision.}
    \label{tab:pipeline-precision-recall}
  \end{subfigure}
   \hspace{0.05\linewidth} 
  \begin{subfigure}[b]{0.45\linewidth}
    \centering
    \includegraphics[width=0.95\linewidth]{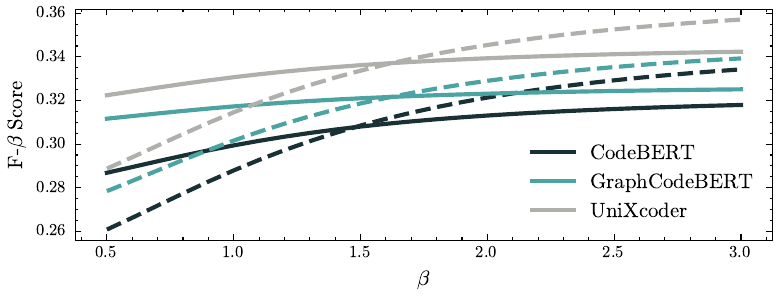}
    \caption{\fbeta{} score for different $\beta$ values for NBDs in \textbf{NN} (solid) and \textbf{Pipeline} (dashed) settings. The \textbf{Pipeline} setting surpasses \textbf{NN} for all NBDs when $\beta\geq1.62$.}
    \label{fig:f-beta}
  \end{subfigure}
  \caption{(a) Precision and recall on the \real{} dataset. (b) F-$\beta{}$ scores for different values of $\beta{}$}
\end{figure*}


\begin{table*}[tb]
    \centering
    \caption{Precision and recall of the bug detectors (a) and type checkers (b) on \annotated{} annotated with types for the NBD and pipeline performance on the unannotated (\emph{raw}) and annotated version. F-scores are omitted since both precision and recall improve consistently when programs are annotated. Numbers are in percentage.}
    \begin{subfigure}[b]{0.7\linewidth}
    \scalebox{0.85}{
    \begin{tabular}{|c|cccc|cccc|}
    \hline
    \multirow{2}{*}{\textbf{Model}} & \multicolumn{4}{c|}{Precision \%} & \multicolumn{4}{c|}{Recall \%} \\
    \cline{2-9}
    & \textbf{NN} & \makecell{\textbf{Pipeline } \\ (\texttt{pytype}-\emph{raw})} & \makecell{\textbf{Pipeline} \\ (\texttt{pytype})} & \makecell{\textbf{Pipeline} \\ (\texttt{mypy})} & \textbf{NN} & \makecell{\textbf{Pipeline} \\ (\texttt{pytype}-\emph{raw})} & \makecell{\textbf{Pipeline} \\ (\texttt{pytype})} & \makecell{\textbf{Pipeline} \\(\texttt{mypy})} \\
    \hline
    CodeBERT      & 74.49 & 72.55 & \textbf{75.65} & 74.79 & 39.25 & 39.78 & 46.77 & \textbf{47.85} \\
    GraphCodeBERT & 72.00 & 69.90 & 73.50 &\textbf{73.77} & 39.78 &  39.78 & 47.51 & \textbf{49.45} \\
    UniXcoder     & 76.24 & 75.00 & \textbf{78.63} & 77.05 & 41.62 & 41.94 & 49.20 & \textbf{50.54} \\
    GGNN          & 46.58 & 46.75 & 59.34 & \textbf{59.79} & 19.88 & 20.93 & 30.68 & \textbf{32.95} \\
    GREAT         & 39.60 & 38.46 & 50.43 & \textbf{51.64} & 24.24 & 24.24 & 34.50 & \textbf{36.84} \\
    \hline
    \end{tabular}
    }
    \caption{}  
    \end{subfigure}
    \begin{subfigure}[b]{0.2\linewidth}
    \scalebox{0.85}{
    \begin{tabular}{|c|cc|}
    \hline
    Type checker & Precision \% & 
    Recall \% \\
    \hline
    \texttt{pytype}-\emph{raw} & 50.00 & 1.50 \\
    \texttt{pytype} & 90.91 & 15.00 \\
    \texttt{mypy} & 87.80 & 18.00 \\
    \hline
    \end{tabular}
    }
    \caption{}  
    \end{subfigure}
    \label{tab:annotated-performance}
\end{table*}

\autoref{tab:pipeline-precision-recall} shows the precision and recall of the NBDs in the \textit{Real} dataset for both the \textbf{NN} and \textbf{Pipeline} settings, and results of the type checker (\texttt{pytype}) on the unannotated \real{} dataset.


The type checker exhibits reasonable precision but low recall in identifying variable misuse bugs when the program is not annotated with types. This limited performance may be attributed to using a single type checker configuration for all programs in the dataset and the lack of type annotation. 


NBDs in the \textbf{Pipeline} configuration demonstrate higher recall than those in the \textbf{NN} setup. Additionally, for certain models (GGNN and GREAT), the precision also improves. 
This is likely caused by the fact that the type checker has better precision than these NBDs.
The recall improvement ranges from $2.31\%$ to $3.97\%$ for all NBDs, indicating that integrating a type checker can help identify more bugs. However, the precision in the \textbf{Pipeline} setting also decreases by between $3.37\%$ to $4.33\%$ for all pre-trained models due to the new false positives introduced by the type checker.


To better understand the trade-off between precision and recall introduced by the type checker on unannotated programs, we plot the \fbeta{} score for different values of $\beta$ in \autoref{fig:f-beta}. The two full training methods are excluded as the \pipeline{} approach improves both precision and recall. In general, when $\beta=1$, the \nn{} setting achieves a higher score compared to the \pipeline{} setting, indicating that it may not be necessary to include type checking when the frequency of false alarms is a concern. However, when $\beta \ge 1.62$ (\textit{i.e.,} recall is considered $1.62$ times more important than precision), \pipeline{} surpasses the \nn{} setting for all NBDs, suggesting that the type checker can be beneficial when recall of bugs is more important.

For the rest of the paper, we consider $\beta=1$ and $\beta=1.5$ on unannotated programs, for cases where precision and recall hold equal importance and cases where 
recall is more critical.

\subsubsection{Results for the \textit{Annotated} dataset}
\autoref{tab:annotated-performance} displays the precision and recall of the bug detectors for both the \textbf{NN} and \textbf{Pipeline} settings, as well as the performance of both type checkers on the \textit{Annotated} dataset. Additionally, we include the performance of these settings on the same dataset without type annotations (\emph{raw}). This comparison is provided because the figures reported for the \textit{Real} dataset in \autoref{tab:pipeline-precision-recall} are not directly comparable due to the \textit{Real} dataset being highly imbalanced, whereas the \textit{Annotated} dataset is balanced.

When the program is annotated with types, the type checker \texttt{pytype} achieves a precision of $90.91\%$ and a recall of $15.00\%$. On unannotated samples, the same type checker achieves a precision of $50\%$ and a recall of $1.50\%$. Conversely, \texttt{mypy}, which only runs on type-annotated samples, achieves a higher recall of $18.00\%$ but a lower precision of $87.80\%$ when compared with \texttt{pytype}. Nevertheless, the recall achieved by type checkers, even with annotations, indicates that these tools fail to detect many variable misuse bugs, highlighting the need to use more advanced tools, such as NBDs, alongside them.

Surprisingly, with the improvement in the performance of the type checker from type annotations, \pipeline{} achieves both higher recall and precision (and thus, higher F-$\beta$ for all $\beta$) compared to \nn{} and \pipeline{} on unannotated programs. On average, the precision of the \pipeline{} setting with \texttt{pytype} is improved by $5.73\%$ while the recall is improved by $8.78\%$  compared to the \nn{} setting. \pipeline{} with \texttt{mypy}  achieves even higher recall (an average increase of $10.57\%$) and a similar improvement in the precision (averaging around $5.63\%$).




\begin{tcolorbox}[left=1mm,right=1mm,top=1mm,bottom=1mm]
    \textbf{RQ2}: Combining a type checker with NBDs can improve recall by 2.31\% to 3.97\%, but may lower precision by $3.37\%$ to $4.33\%$ on unannotated programs. On annotated programs, integrating a type checker improves both precision and recall. Thus, the pipeline setting is beneficial when recall is key or when type annotations are present.
    
    
\end{tcolorbox}

\subsection{RQ3: Evaluation on type-related bugs}
\subsubsection{Addressing the RQ}
Since integrating type checkers can be beneficial if identifying bugs is key or when type annotations exist, when using a type checker in combination with an NBD, the NBD should mainly focus on the bugs that are not easily detectable by type checkers. In this section, we evaluate the performance of NBDs on type-related bugs compared to other bugs when trained with a synthetic dataset on the variable misuse task. We divide the evaluation dataset into two groups based on the type-checking results in the evaluation: type-related bugs and other bugs. 
We then show the impact of the type-related bugs on overall performance by evaluating all programs and programs filtering type-related bugs. In this experiment, we consider \synthetic{}, \real{}, and the \annotated{} datasets and filter the bugs with \texttt{pytype}.

\begin{table*}[tb]
    \centering
    \caption{Impact of filtering type-related bugs on precision and recall. Numbers are in percentage. F-scores are omitted since both precision and recall decrease consistently.
    }
    
\scalebox{0.8}{
    \begin{tabular}{|c|ccc|ccc||ccc|ccc||ccc|ccc|}
    \hline
    \multirow{3}{*}{\textbf{Model}} &\multicolumn{6}{c||}{\synthetic{}} & \multicolumn{6}{c||}{\real{}} & \multicolumn{6}{c|}{\textit{Annotated}}\\
    \cline{2-19}
    & \multicolumn{3}{c|}{Precision \%} & \multicolumn{3}{c||}{Recall \%} & \multicolumn{3}{c|}{Precision \%} & \multicolumn{3}{c||}{Recall \%} & \multicolumn{3}{c|}{Precision \%} & \multicolumn{3}{c|}{Recall \%} \\
    \cline{2-19}
    & \textbf{Full} & \textbf{Filtered} & $\Delta$ & \textbf{Full} & \textbf{Filtered} & $\Delta$ & \textbf{Full} & \textbf{Filtered} &$\Delta$ & \textbf{Full} & \textbf{Filtered} &$\Delta$ & \textbf{Full} & \textbf{Filtered} &$\Delta$ & \textbf{Full} & \textbf{Filtered} &$\Delta$\\
    \hline
    CodeBERT      & 93.40 & 91.91 & -1.59 & 91.67 & 90.77 & -0.98 & 27.91 & 26.10 & -6.48 &   32.29 & 31.02 & -3.95 & 74.49 & 69.51 & -6.68 & 39.25 & 36.54 & -6.90    \\
    GraphCodeBERT & 94.00 & 92.63 & -1.46 & 92.19 & 91.38 & -0.88 & 30.80 & 28.80 & -6.49 &   32.71 & 31.27 & -4.42 & 72.00 & 66.67 & -7.41 & 39.78 & 37.09 & -6.77    \\
    UniXcoder     & 94.07 & 92.70 & -1.45 & 93.58 & 92.79 & -0.84 & 31.71 & 29.92 & -5.67 &   34.54 & 33.32 & -3.55 & 76.24 & 73.81 & -3.18 & 41.62 & 39.49 & -5.12   \\
    GGNN          & 86.38 & 83.12 & -3.77 & 80.89 & 77.31 & -4.42 & 10.51 & 9.47  & -9.88 & 17.47   & 16.43 & -5.93 & 46.58 & 41.38 & -11.16 & 19.88 & 16.44 & -17.32   \\
    GREAT         & 84.80 & 81.44 & -3.97 & 89.68 & 87.53 & -2.39 & 9.86  & 8.84  & -10.35 & 26.59   & 24.95 & -6.16 & 39.60 & 34.52 & -12.83 & 24.24 & 20.57 & -15.16    \\
    \hline
    \end{tabular}    
}
    \label{tab:type-impact}
\end{table*}

 

\subsubsection{Results}

\autoref{tab:type-impact} shows the precision and recall of different NBDs.
Comparing the performance of the NBDs on synthetic and real-world datasets, we observe that these detectors perform much better on the synthetic datasets compared to the real-world bugs, consistent with previous work~\cite{he2022distribution,allamanis2021self}. This discrepancy can be attributed to the distribution shift between synthetic and real-world bugs~\cite{he2022distribution}. 



To evaluate the impact of type-related bugs on precision and recall, we include the ratio of change ($\Delta$) in the filtered dataset compared to the full dataset. We observe a decrease in both precision and recall for all NBDs in the \emph{filtered} evaluation sets, hinting at a bias of NBDs towards type-related bugs. 

For unannotated programs, removing type-related bugs from the evaluation dataset leads to an average reduction in precision ($5.11\%$) and recall ($3.35\%$). The NBDs exhibit a more notable decrease in \real{} compared to \synthetic{}, with an average of $7.77\%$ and $4.80\%$ for precision and recall, respectively.  
With type annotations (\annotated{}), the decrease is even more notable, with the precision of NBDs dropping by $3.18\%$ to $12.83\%$ and recall dropping by $5.12\%$ to $17.32\%$.

These findings imply that type-related bugs in the evaluation may lead to overly optimistic outcomes, especially when a type checker is used in the pipeline. Furthermore, the decline in performance is notably more significant for the real-world datasets compared to the synthetic dataset. This observation brings increased threats to evaluation validity in real-world use cases compared to the synthetic dataset.

\begin{tcolorbox}[left=1mm,right=1mm,top=1mm,bottom=1mm]
\textbf{RQ3}: Our findings reveal that bug detectors exhibit superior performance in detecting type-related bugs compared to other types of bugs. Removing these bugs from the evaluation dataset leads to a reduction in both precision (an average of 5.11\%) and recall (an average of 3.35\%) even in the absence of type annotations. When type annotation exists, the dropping in performance is even more significant. This suggests that the inclusion of type-related bugs in the evaluation dataset, particularly when using a type checker, may yield overly optimistic evaluation results.

\end{tcolorbox}

\subsection{RQ4: Effects of type-related bugs on training}

\subsubsection{Addressing the RQ}
In this section, we investigate the impact of filtering type-related bugs from the training data on the performance of the NBDs. NBDs trained with the \emph{full dataset} are compared with those trained with the \emph{filtered dataset} using precision, recall, and \fbeta{} on all samples from the evaluation datasets, excluding type-related bugs. We aim to measure the performance of NBDs when the type checker is used together as a pipeline. Thus, we only focus on bugs beyond the scope of type checkers and filter all type-related bugs from the datasets. Since manually annotating the training dataset would be overwhelming due to the large size, the training dataset is filtered using unannotated programs only. To ensure consistency between training and evaluation, we exclude the \annotated{} subset in this experiment and leave the exploration of type annotations in \emph{training} as future work.

A naive approach to filter type-related bugs from the training data is to remove all samples identified as type-related bugs. However, this approach may result in a significant reduction in the size of the training data as well as in an imbalance in the correct and faulty programs. To mitigate these issues, we \textit{replace} the type-related bugs with other bugs in the training data by randomly oversampling other bugs. This approach ensures that the size of the training data remains the same and the balance between correct and faulty programs is maintained.


\begin{table*}[tb]
    \centering
    \caption{Performance of detectors trained with \textit{full dataset} $\rightarrow$ \textit{filtered dataset}. Numbers are in percentage. In general, filtering type-related bugs in training improves the recall while decreasing the precision.}
    \scalebox{0.95}{
    \begin{tabular}{|c|cccc|cccc|}
    \hline
    \multirow{2}{*}{\textbf{Model}} & \multicolumn{4}{c|}{\synthetic{}} & \multicolumn{4}{c|}{\textit{Real}} \\
    \cline{2-9}
    & \textbf{Precision \%} & \textbf{Recall \%} & F-1 \% & \fbeta{} ($\beta=1.5$) \%& \textbf{Precision \%} & \textbf{Recall \%} & F-1 \% & \fbeta{} ($\beta=1.5$) \%\\
    \hline
    CodeBERT      & 91.91$\rightarrow$91.38 &  90.77$\rightarrow$91.17 & \textbf{91.33}$\rightarrow$91.28 & 91.12$\rightarrow$\textbf{91.23} &  26.10$\rightarrow$24.77 &  31.02$\rightarrow$31.06 & \textbf{28.35}$\rightarrow$27.56 & \textbf{29.32}$\rightarrow$28.81 \\
    GraphCodeBERT & 92.63$\rightarrow$92.12 &  91.38$\rightarrow$91.65 & \textbf{92.00}$\rightarrow$91.88 & 91.76$\rightarrow$\textbf{91.79} &  28.80$\rightarrow$28.69 &  31.27$\rightarrow$34.35 & 29.99$\rightarrow$\textbf{31.26} & 30.47$\rightarrow$\textbf{32.38} \\
    UniXcoder     & 92.70$\rightarrow$92.22 &  92.79$\rightarrow$93.10 & \textbf{92.75}$\rightarrow$92.66 & 92.76$\rightarrow$\textbf{92.83} &  29.92$\rightarrow$28.89 &  33.32$\rightarrow$35.98 & 31.52$\rightarrow$\textbf{32.05} & 32.19$\rightarrow$\textbf{33.45} \\
    GGNN          & 83.12$\rightarrow$82.78 &  77.31$\rightarrow$78.18 & 80.11$\rightarrow$\textbf{80.42} & 79.01$\rightarrow$\textbf{79.54} &   9.47$\rightarrow$9.33  &  16.43$\rightarrow$17.04 & 12.02$\rightarrow$\textbf{12.06} & 13.40$\rightarrow$\textbf{13.58} \\
    GREAT         & 81.44$\rightarrow$82.20 &  87.53$\rightarrow$87.72 & 84.38$\rightarrow$\textbf{84.87} & 85.56$\rightarrow$\textbf{85.95} &   8.84$\rightarrow$10.06 &  24.95$\rightarrow$27.12 & 13.05$\rightarrow$\textbf{14.67} & 15.98$\rightarrow$\textbf{17.82} \\
    \hline
    \end{tabular}
    }
    \label{tab:impact-training} 
\end{table*}

    

\subsubsection{Results}

The performance change of NBDs trained with the full dataset and the filtered dataset is shown in \autoref{tab:impact-training}. Values on the left side of the arrow show the performance of the full training dataset while the right side represents the performance of filtered training. The better \fbeta{} scores of the two settings are highlighted in bold.


In general, the recall is improved when the NBD is trained on the filtered dataset (observed in all cases) with an average improvement of $1.06\%$.
This improvement indicates that the removal of type-related bugs from the training data enables the NBDs to focus more on other non-type-related bugs. On the other hand, the precision is also reduced in 8 out of 10 cases. However, the change is generally smaller, with an average decrease of $0.56\%$.
We hypothesize that this decrease may be attributed to the oversampling strategy we used, rather than generating entirely new synthetic samples.

When assessing the \fbeta{} score, we evaluate both cases of $\beta=1$ (F-1) and $\beta=1.5$. When using the F-1 score, where recall is equally important as precision, filtering out type-related bugs from the training data does not yield a visible improvement in overall performance, as indicated by a better F-1 score in only 6 out of 10 cases. This finding aligns with our observation in \autoref{sec:influence-type-check} that integrating type checkers may not be beneficial in these cases.

When recall is more important
($\beta=1.5$), we note that the filtered dataset consistently outperforms the full dataset in all cases for the synthetic dataset. This observation suggests that the reduction in precision is less pronounced compared to the increase in recall 
in this case. This improvement extends to the real-world dataset as well, with the filtered dataset outperforming the full dataset in 4 out of 5 cases while not outperforming in the case of CodeBERT. The slight variation in results for the real-world dataset may stem from the distribution shift between synthetic and real-world bugs.

\begin{tcolorbox}[left=1mm,right=1mm,top=1mm,bottom=1mm]
    \textbf{RQ4}: Removing type-related errors from the training dataset causes a significant increase in recall. Although a decrease in precision is also noted, the overall \fbeta{} score is improved in 9 out of 10 cases when $\beta=1.5$. 
    
\end{tcolorbox}
\section{Discussion}
\label{sec:discusssion}

\subsection{Key implications}
We show that datasets for Python variable misuse contain a significant portion of type-related bugs. Moreover, we conduct systematic experiments to evaluate the impact of type-related bugs on the training and evaluation of NBDs. Based on the results, we provide practical takeaways for using NBDs \emph{in detecting variable misuse bugs} for dynamically typed languages. 

\begin{tcolorbox}[left=1mm,right=1mm,top=1mm,bottom=1mm]
\textbf{Takeaway 1}:
 In scenarios where developers include type annotations, type checkers should be used in conjunction with NBDs. If no annotations are provided, combining type checkers with NBDs is beneficial when prioritizing recall.
\end{tcolorbox}

Many existing studies have demonstrated that type checkers can help identify bugs in dynamically typed languages when manual type annotation is provided \cite{khan2021empirical,gao2017type}. Our study builds upon these findings, revealing that type checkers are capable of detecting a significant proportion of real-world bugs, around 4.95\%, even in the absence of explicit type annotations. If the programs are annotated, the percentage of type-related bugs detected increases to a range of 15.00\% to 18.00\%, depending on the type checker used.

Based on these observations, we assess the combined use of type checkers and NBDs on both unannotated and annotated programs. Analysis of this assessment indicates that 
if recall is more important than precision (\textit{i.e.,} when $\beta \ge 1.62$) including type checkers is recommended. When type annotations exist, integrating type checkers is always beneficial.




\begin{tcolorbox}[left=1mm,right=1mm,top=1mm,bottom=1mm]
\textbf{Takeaway 2}: When evaluating NBDs, type-related bugs and other bugs should be evaluated separately. 
\end{tcolorbox}

Existing assessments of NBDs typically combine all bug types into a unified dataset. Our findings reveal that excluding type-related bugs from the evaluation dataset significantly reduces the NBDs' performance. Thus, to accurately estimate the effectiveness of NBDs in conjunction with a type checker, removal of these type-related bugs from the evaluation is necessary. Conversely, if the aim is to evaluate the NBD's stand-alone performance, a more detailed bug categorization should be used to comprehensively understand their capabilities.




\begin{tcolorbox}[left=1mm,right=1mm,top=1mm,bottom=1mm]
\textbf{Takeaway 3}: When training an NBD to complement type checkers, eliminating type-related bugs from the training datasets enhances the neural model's ability to detect non-type-related bugs. However, this refinement often leads to a decrease in precision.

\end{tcolorbox}
Excluding type-related bugs from unannotated datasets during the training phase results in general improvements in recall for all NBDs within both synthetic and real-world datasets. However, this improvement in recall comes at the cost of a reduction in precision. Thus, the decision to exclude these bugs from training depends on the specific goal of the bug detector. If recall is more important than precision,
then eliminating type-related bugs during training is advantageous.




\subsection{Threats to validity}
\paragraph{Internal validity}
Training and fine-tuning of NBDs can be sensitive to the choice of hyperparameters. We mitigate this issue by reusing hyperparameters from previous work \cite{hellendoorn2019global}. Furthermore, we use the same hyperparameters for the same model architecture when training with the full and filtered dataset. Moreover, the training process can be non-deterministic and, due to the cost of training, we only used one random seed for each setting. To mitigate this issue, we experimented with multiple different neural network architectures and training paradigms. 
When evaluating the impact of type-related bugs in the training data, we oversample bugs to keep the dataset balanced. Such a strategy provides significant improvement in recall. Another approach to further improve performance (but left for future work) could be to generate more synthetic bugs while filtering type-related bugs. 
We manually annotated a subset of the \real{} dataset to evaluate the performance when type annotation exists. To ensure accuracy and minimize bias, we only use the correct version of the programs and follow procedures from previous work.

\paragraph{External validity}
This paper exclusively concentrates on Python due to the limited availability of variable misuse datasets for other dynamically typed languages. Consequently, the insights drawn from this study might not be directly transferable to other dynamically typed languages. We defer the exploration of such intriguing avenues to future research.
The impact of type checking may differ for NBDs. We study various recent NBDs and obtained similar effects. Besides, we rely solely on type checkers to detect and classify type-related bugs due to the lack of ground truth. Implementation errors in the type checkers may cause incorrect detection. To mitigate this issue, we use popular type checkers developed and maintained by Google and the Python community. 


\paragraph{Construct validity}
Selecting appropriate performance metrics to evaluate performance in the variable misuse task presents a potential threat to validity. To tackle this concern, we use standard metrics for classification by considering bug detection as a classification problem. These metrics resemble existing metrics used for NBDs. In assessing the joint effectiveness of NBDs, we employed the \fbeta{} score. The choice of $\beta$ values is tailored to balance the emphasis between identifying bugs and minimizing false alarms. To address variability in performance, we present results across a spectrum of $\beta$ values during pipeline performance evaluation. Furthermore, we select two $\beta$ values to represent different scenarios in our analysis of the influence of type-related bugs on training.

\section{Related Work}
\label{sec:rel-work}

\subsection{Neural bug detectors}

\paragraph{Target bug types}

Multiple studies have shown the effectiveness of using neural networks for bug detection. Besides variable misuse bugs studied in this paper, neural networks have been used to detect other types of bugs, such as operand swapping \cite{troshin2022probing}, wrong binary operator \cite{troshin2022probing,he2022distribution,allamanis2021self}, and identifying wrong argument or return values of a function \cite{allamanis2021self,li2020deep}. Most of the work adapts the strategy of joint learning of detection and repair of the program together~\cite{VasicKMBS19}.

\paragraph{Proposed models}

The predominant models utilized for bug detection can be broadly categorized into Transformers, GNNs, and hybrid methods~\cite{hellendoorn2019global}. Transformers process programs as sequences of tokens, with the flexibility of being either trained from scratch~\cite{hellendoorn2019global} or fine-tuned (\textit{e.g.,} CuBERT~\cite{kanade2020learning}, CodeBERT~\cite{feng2020codebert}, GraphCodeBERT~\cite{guo2020graphcodebert}, UniXcoder~\cite{guo2022unixcoder}, etc.). GNN models, on the other hand, interpret programs in the form of graphs featuring diverse edges such as control flow, data flow, etc.~\cite{allamanis2017learning}. Hybrid approaches combine both models. For example, GREAT~\cite{hellendoorn2019global} combines the transformer architecture with edge types from the program graph. The sandwich model~\cite{hellendoorn2019global} alternates between transformer layers and graph neural layers, forming a cohesive approach.

This paper explores a range of models including CodeBERT, GraphCodeBERT, UniXcoder, GGNN, and GREAT, thereby encompassing the entire spectrum of commonly employed approaches for bug detection.

\paragraph{Datasets}
Due to the lack of real-world bugs, training on real-world bugs alone is typically not sufficient to obtain a high-performing bug detector \cite{he2022distribution,richter2023train}. As a result, many approaches have been focused on generating synthetic bugs. The most common approach is to randomly perturb a program with a set of predefined mutation rules \cite{allamanis2017learning,VasicKMBS19}.

Recent work demonstrates that neural detectors trained on synthetic bugs do not perform well on real-world bugs \cite{allamanis2022graph}. This difference in performance can be attributed to the different characteristics between synthetic and real-world bugs \cite{alrashedy2023learning,he2022distribution}. As a result, many studies have experimented with mixing synthetic and real-world bugs in training \cite{he2022distribution,richter2023train}.

At the same time, other approaches attempt to generate synthetic datasets that are more similar to real-world bugs. For example, Richter et al. \cite{richter2022learning} use a learning based approach to generate faulty programs taking the context of the original program into account. SemSeed \cite{patra2021semantic} extracts a set of bug patterns from real-world bugs and applies the pattern most similar to the original program to generate a synthetic bug.

Our study takes a different perspective on analyzing the datasets for training NBDs. We investigate how type-checking affects the performance of NBDs and offer recommendations to enhance their practicality when deployed in such scenarios.



\subsection{Combining static analysis and neural networks}
Static analysis has long been used to enhance the performance of neural networks used on code. In NBDs, static analysis generates a program graph used as input to the neural network \cite{allamanis2017learning,hellendoorn2019global}. When repairing the bug, feedback generated from the compiler is used as additional information to help the neural network \cite{yasunaga2020graph,ye2022selfapr,tarlow2020learning}.

Another task commonly associated with static analysis is type inference. When employing a neural network for variable type prediction, particularly in dynamically typed languages, the predicted types often lack guarantees of consistency with the type checker \cite{cassano2023type,allamanis2020typilus,yee2023machine}. In such scenarios, a type checker is used to (1) guide the search space of types and (2) ensure the coherence of the predicted types \cite{pradel2020typewriter,cassano2023type,wei2022typet5,pandi2020opttyper}.

Our work shows how and when a type checker should be used together with NBDs to improve the overall bug detection performance. We also show how bugs detectable by a type checker can influence the training and evaluation of NBDs.

\section{Conclusion and future work}
\label{sec:conclusion}

In this study, we explore the impact of type checking on automated bug detection using neural networks, focusing on variable misuse errors. We found that over 19\% of synthetic bugs and around 5\% of real-world bugs in datasets used for NBDs can be identified by a type checker without explicit type annotations. When type annotaion are present, over 15\% of real-world bugs can be detected. Via experiments with various NBDs, we find that the benefits of type-checking are most pronounced 
when recall is more important than precision or when type annotations exist. When the NBDs are used together with a type checker, it has been observed that type-related bugs lead to an overly optimistic evaluation. We also discover that excluding these bugs from training can enhance the detection capabilities of NBDs while reducing their precision. As a result, the decision to exclude type-related bugs from training depends on whether the priority is recall or precision 

We plan to extend the work from three aspects. First, the impact of type checking on dynamically typed languages beyond Python and other types of bugs can be investigated. 
Then, the impact of different bug types in type-related errors on NBDs can be studied. Finally, the influence of neural type inference can be explored.


\bibliographystyle{IEEEtran}
\bibliography{reference}

\begin{thebibliography}{10}
\providecommand{\url}[1]{#1}
\csname url@samestyle\endcsname
\providecommand{\newblock}{\relax}
\providecommand{\bibinfo}[2]{#2}
\providecommand{\BIBentrySTDinterwordspacing}{\spaceskip=0pt\relax}
\providecommand{\BIBentryALTinterwordstretchfactor}{4}
\providecommand{\BIBentryALTinterwordspacing}{\spaceskip=\fontdimen2\font plus
\BIBentryALTinterwordstretchfactor\fontdimen3\font minus \fontdimen4\font\relax}
\providecommand{\BIBforeignlanguage}[2]{{%
\expandafter\ifx\csname l@#1\endcsname\relax
\typeout{** WARNING: IEEEtran.bst: No hyphenation pattern has been}%
\typeout{** loaded for the language `#1'. Using the pattern for}%
\typeout{** the default language instead.}%
\else
\language=\csname l@#1\endcsname
\fi
#2}}
\providecommand{\BIBdecl}{\relax}
\BIBdecl

\bibitem{mcconnell2004code}
S.~McConnell, \emph{Code complete - A practical handbook of software construction, 2nd Edition}.\hskip 1em plus 0.5em minus 0.4em\relax Microsoft Press, 2004.

\bibitem{habib2018many}
A.~Habib and M.~Pradel, ``How many of all bugs do we find? {A} study of static bug detectors,'' in \emph{Proceedings of the 33rd {ACM/IEEE} International Conference on Automated Software Engineering, {ASE} 2018, Montpellier, France, September 3-7, 2018}.\hskip 1em plus 0.5em minus 0.4em\relax {ACM}, 2018, pp. 317--328.

\bibitem{fonseca2017empirical}
P.~Fonseca, K.~Zhang, X.~Wang, and A.~Krishnamurthy, ``An empirical study on the correctness of formally verified distributed systems,'' in \emph{Proceedings of the Twelfth European Conference on Computer Systems, EuroSys 2017, Belgrade, Serbia, April 23-26, 2017}.\hskip 1em plus 0.5em minus 0.4em\relax {ACM}, 2017, pp. 328--343.

\bibitem{jemal2022presence}
S.~Jemal, ``On the presence and causes of lingering defects in software: An industrial study of lingering defects,'' Ph.D. dissertation, Blekinge Institute of Technology, 2022.

\bibitem{poulsen2004software}
\BIBentryALTinterwordspacing
K.~Poulsen, ``Software bug contributed to blackout,'' 2004. [Online]. Available: \url{https://www.theregister.com/2004/02/12/software_bug_contributed_to_blackout/}
\BIBentrySTDinterwordspacing

\bibitem{zhivich2009real}
M.~Zhivich and R.~K. Cunningham, ``The real cost of software errors,'' \emph{{IEEE} Secur. Priv.}, vol.~7, no.~2, pp. 87--90, 2009.

\bibitem{ray2014large}
B.~Ray, D.~Posnett, P.~T. Devanbu, and V.~Filkov, ``A large-scale study of programming languages and code quality in {GitHub},'' \emph{Commun. {ACM}}, vol.~60, no.~10, pp. 91--100, 2017.

\bibitem{gao2017type}
Z.~Gao, C.~Bird, and E.~T. Barr, ``To type or not to type: Quantifying detectable bugs in {JavaScript},'' in \emph{Proceedings of the 39th International Conference on Software Engineering, {ICSE} 2017, Buenos Aires, Argentina, May 20-28, 2017}.\hskip 1em plus 0.5em minus 0.4em\relax {IEEE} / {ACM}, 2017, pp. 758--769.

\bibitem{khan2021empirical}
F.~Khan, B.~Chen, D.~Varr{\'{o}}, and S.~McIntosh, ``An empirical study of type-related defects in {Python} projects,'' \emph{{IEEE} Trans. Software Eng.}, vol.~48, no.~8, pp. 3145--3158, 2022.

\bibitem{aftandilian2012building}
E.~Aftandilian, R.~Sauciuc, S.~Priya, and S.~Krishnan, ``Building useful program analysis tools using an extensible {Java} compiler,'' in \emph{12th {IEEE} International Working Conference on Source Code Analysis and Manipulation, {SCAM} 2012, Riva del Garda, Italy, September 23-24, 2012}.\hskip 1em plus 0.5em minus 0.4em\relax {IEEE} Computer Society, 2012, pp. 14--23.

\bibitem{calcagno2015moving}
C.~Calcagno, D.~Distefano, J.~Dubreil, D.~Gabi, P.~Hooimeijer, M.~Luca, P.~W. O'Hearn, I.~Papakonstantinou, J.~Purbrick, and D.~Rodriguez, ``Moving fast with software verification,'' in \emph{{NASA} Formal Methods - 7th International Symposium, {NFM} 2015, Pasadena, CA, USA, April 27-29, 2015, Proceedings}, ser. Lecture Notes in Computer Science, vol. 9058.\hskip 1em plus 0.5em minus 0.4em\relax Springer, 2015, pp. 3--11.

\bibitem{hovemeyer2004finding}
D.~Hovemeyer and W.~W. Pugh, ``Finding bugs is easy,'' \emph{{ACM} {SIGPLAN} Notices}, vol.~39, no.~12, pp. 92--106, 2004.

\bibitem{habib2019neural}
A.~Habib and M.~Pradel, ``Neural bug finding: {A} study of opportunities and challenges,'' \emph{CoRR}, vol. abs/1906.00307, 2019.

\bibitem{allamanis2017learning}
M.~Allamanis, M.~Brockschmidt, and M.~Khademi, ``Learning to represent programs with graphs,'' in \emph{6th International Conference on Learning Representations, {ICLR} 2018, Vancouver, BC, Canada, April 30 - May 3, 2018, Conference Track Proceedings}, 2018.

\bibitem{troshin2022probing}
S.~Troshin and N.~Chirkova, ``Probing pretrained models of source codes,'' in \emph{Proceedings of the Fifth BlackboxNLP Workshop on Analyzing and Interpreting Neural Networks for NLP, BlackboxNLP@EMNLP 2022, Abu Dhabi, United Arab Emirates (Hybrid), December 8, 2022}.\hskip 1em plus 0.5em minus 0.4em\relax Association for Computational Linguistics, 2022, pp. 371--383.

\bibitem{VasicKMBS19}
M.~Vasic, A.~Kanade, P.~Maniatis, D.~Bieber, and R.~Singh, ``Neural program repair by jointly learning to localize and repair,'' in \emph{7th International Conference on Learning Representations, {ICLR} 2019, New Orleans, LA, USA, May 6-9, 2019}, 2019.

\bibitem{kanade2020learning}
A.~Kanade, P.~Maniatis, G.~Balakrishnan, and K.~Shi, ``Learning and evaluating contextual embedding of source code,'' in \emph{Proceedings of the 37th International Conference on Machine Learning, {ICML} 2020, 13-18 July 2020, Virtual Event}, ser. Proceedings of Machine Learning Research, vol. 119.\hskip 1em plus 0.5em minus 0.4em\relax {PMLR}, 2020, pp. 5110--5121.

\bibitem{hellendoorn2019global}
V.~J. Hellendoorn, C.~Sutton, R.~Singh, P.~Maniatis, and D.~Bieber, ``Global relational models of source code,'' in \emph{8th International Conference on Learning Representations, {ICLR} 2020, Addis Ababa, Ethiopia, April 26-30, 2020}, 2020.

\bibitem{di2022evolution}
L.~D. Grazia and M.~Pradel, ``The evolution of type annotations in {Python}: An empirical study,'' in \emph{Proceedings of the 30th {ACM} Joint European Software Engineering Conference and Symposium on the Foundations of Software Engineering, {ESEC/FSE} 2022, Singapore, Singapore, November 14-18, 2022}.\hskip 1em plus 0.5em minus 0.4em\relax {ACM}, 2022, pp. 209--220.

\bibitem{li2015gated}
Y.~Li, D.~Tarlow, M.~Brockschmidt, and R.~S. Zemel, ``Gated graph sequence neural networks,'' in \emph{4th International Conference on Learning Representations, {ICLR} 2016, San Juan, Puerto Rico, May 2-4, 2016, Conference Track Proceedings}, 2016.

\bibitem{feng2020codebert}
Z.~Feng, D.~Guo, D.~Tang, N.~Duan, X.~Feng, M.~Gong, L.~Shou, B.~Qin, T.~Liu, D.~Jiang, and M.~Zhou, ``{CodeBERT}: {A} pre-trained model for programming and natural languages,'' in \emph{Findings of the Association for Computational Linguistics: {EMNLP} 2020, Online Event, 16-20 November 2020}, ser. Findings of {ACL}, vol. {EMNLP} 2020.\hskip 1em plus 0.5em minus 0.4em\relax Association for Computational Linguistics, 2020, pp. 1536--1547.

\bibitem{guo2020graphcodebert}
D.~Guo, S.~Ren, S.~Lu, Z.~Feng, D.~Tang, S.~Liu, L.~Zhou, N.~Duan, A.~Svyatkovskiy, S.~Fu, M.~Tufano, S.~K. Deng, C.~B. Clement, D.~Drain, N.~Sundaresan, J.~Yin, D.~Jiang, and M.~Zhou, ``{GraphCodeBERT}: Pre-training code representations with data flow,'' in \emph{9th International Conference on Learning Representations, {ICLR} 2021, Virtual Event, Austria, May 3-7, 2021}, 2021.

\bibitem{guo2022unixcoder}
D.~Guo, S.~Lu, N.~Duan, Y.~Wang, M.~Zhou, and J.~Yin, ``{UniXcoder}: Unified cross-modal pre-training for code representation,'' in \emph{Proceedings of the 60th Annual Meeting of the Association for Computational Linguistics (Volume 1: Long Papers), {ACL} 2022, Dublin, Ireland, May 22-27, 2022}.\hskip 1em plus 0.5em minus 0.4em\relax Association for Computational Linguistics, 2022, pp. 7212--7225.

\bibitem{rabin2021understanding}
M.~R.~I. Rabin, V.~J. Hellendoorn, and M.~A. Alipour, ``Understanding neural code intelligence through program simplification,'' in \emph{Proceedings of the 29th ACM Joint Meeting on European Software Engineering Conference and Symposium on the Foundations of Software Engineering}, 2021, pp. 441--452.

\bibitem{richter2023train}
C.~Richter and H.~Wehrheim, ``How to train your neural bug detector: Artificial vs real bugs,'' in \emph{38th {IEEE/ACM} International Conference on Automated Software Engineering, {ASE} 2023, Luxembourg, September 11-15, 2023}.\hskip 1em plus 0.5em minus 0.4em\relax {IEEE}, 2023, pp. 1036--1048.

\bibitem{tarlow2020learning}
D.~Tarlow, S.~Moitra, A.~Rice, Z.~Chen, P.~Manzagol, C.~Sutton, and E.~Aftandilian, ``Learning to fix build errors with {Graph2Diff} neural networks,'' in \emph{{ICSE} '20: 42nd International Conference on Software Engineering, Workshops, Seoul, Republic of Korea, 27 June - 19 July, 2020}.\hskip 1em plus 0.5em minus 0.4em\relax {ACM}, 2020, pp. 19--20.

\bibitem{karampatsis2020often}
R.~Karampatsis and C.~Sutton, ``How often do single-statement bugs occur?: The {ManySStuBs4J} dataset,'' in \emph{{MSR} '20: 17th International Conference on Mining Software Repositories, Seoul, Republic of Korea, 29-30 June, 2020}.\hskip 1em plus 0.5em minus 0.4em\relax {ACM}, 2020, pp. 573--577.

\bibitem{allamanis2022graph}
M.~Allamanis, ``Graph neural networks in program analysis,'' in \emph{Graph neural networks: foundations, frontiers, and applications}.\hskip 1em plus 0.5em minus 0.4em\relax Springer, 2022, pp. 483--497.

\bibitem{artifacts_icse25}
\BIBentryALTinterwordspacing
B.~Chen, J.~A. Hern\'andez~L\'opez, G.~Mussbacher, and D.~Varr\'o, ``Paper artifacts.'' [Online]. Available: \url{https://github.com/20001LastOrder/icse2025-type4py}
\BIBentrySTDinterwordspacing

\bibitem{rak2020python}
I.~Rak{-}amnouykit, D.~McCrevan, A.~L. Milanova, M.~Hirzel, and J.~Dolby, ``Python 3 types in the wild: a tale of two type systems,'' in \emph{{DLS} 2020: Proceedings of the 16th {ACM} {SIGPLAN} International Symposium on Dynamic Languages, Virtual Event, USA, November 17, 2020}.\hskip 1em plus 0.5em minus 0.4em\relax {ACM}, 2020, pp. 57--70.

\bibitem{PEP484}
\BIBentryALTinterwordspacing
G.~van Rossum, J.~Lehtosalo, and L.~Langa, ``Pep 484 – type hints.'' [Online]. Available: \url{https://peps.python.org/pep-0484/}
\BIBentrySTDinterwordspacing

\bibitem{mypy}
\BIBentryALTinterwordspacing
``mypy - optional static typing for {Python}.'' [Online]. Available: \url{https://mypy-lang.org/}
\BIBentrySTDinterwordspacing

\bibitem{siek2007gradual}
J.~G. Siek and W.~Taha, ``Gradual typing for objects,'' in \emph{{ECOOP} 2007 - Object-Oriented Programming, 21st European Conference, Berlin, Germany, July 30 - August 3, 2007, Proceedings}, ser. Lecture Notes in Computer Science, vol. 4609.\hskip 1em plus 0.5em minus 0.4em\relax Springer, 2007, pp. 2--27.

\bibitem{pytype}
\BIBentryALTinterwordspacing
``pytype - a static type analyzer for {Python} code.'' [Online]. Available: \url{https://google.github.io/pytype/}
\BIBentrySTDinterwordspacing

\bibitem{hellendoorn2018deep}
V.~J. Hellendoorn, C.~Bird, E.~T. Barr, and M.~Allamanis, ``Deep learning type inference,'' in \emph{Proceedings of the 2018 {ACM} Joint Meeting on European Software Engineering Conference and Symposium on the Foundations of Software Engineering, {ESEC/SIGSOFT} {FSE} 2018, Lake Buena Vista, FL, USA, November 04-09, 2018}.\hskip 1em plus 0.5em minus 0.4em\relax {ACM}, 2018, pp. 152--162.

\bibitem{allamanis2020typilus}
M.~Allamanis, E.~T. Barr, S.~Ducousso, and Z.~Gao, ``Typilus: Neural type hints,'' in \emph{Proceedings of the 41st {ACM} {SIGPLAN} International Conference on Programming Language Design and Implementation, {PLDI} 2020, London, UK, June 15-20, 2020}.\hskip 1em plus 0.5em minus 0.4em\relax {ACM}, 2020, pp. 91--105.

\bibitem{vaswani2017attention}
A.~Vaswani, N.~Shazeer, N.~Parmar, J.~Uszkoreit, L.~Jones, A.~N. Gomez, L.~Kaiser, and I.~Polosukhin, ``Attention is all you need,'' in \emph{Advances in Neural Information Processing Systems 30: Annual Conference on Neural Information Processing Systems 2017, December 4-9, 2017, Long Beach, CA, {USA}}, 2017, pp. 5998--6008.

\bibitem{lin2022survey}
T.~Lin, Y.~Wang, X.~Liu, and X.~Qiu, ``A survey of transformers,'' \emph{{AI} Open}, vol.~3, pp. 111--132, 2022.

\bibitem{kipf2016semi}
T.~N. Kipf and M.~Welling, ``Semi-supervised classification with graph convolutional networks,'' in \emph{5th International Conference on Learning Representations, {ICLR} 2017, Toulon, France, April 24-26, 2017, Conference Track Proceedings}, 2017.

\bibitem{cho2014properties}
K.~Cho, B.~van Merrienboer, D.~Bahdanau, and Y.~Bengio, ``On the properties of neural machine translation: Encoder-decoder approaches,'' in \emph{Proceedings of SSST@EMNLP 2014, Eighth Workshop on Syntax, Semantics and Structure in Statistical Translation, Doha, Qatar, 25 October 2014}.\hskip 1em plus 0.5em minus 0.4em\relax Association for Computational Linguistics, 2014, pp. 103--111.

\bibitem{he2022distribution}
J.~He, L.~Beurer{-}Kellner, and M.~T. Vechev, ``On distribution shift in learning-based bug detectors,'' in \emph{International Conference on Machine Learning, {ICML} 2022, 17-23 July 2022, Baltimore, Maryland, {USA}}, ser. Proceedings of Machine Learning Research, vol. 162.\hskip 1em plus 0.5em minus 0.4em\relax {PMLR}, 2022, pp. 8559--8580.

\bibitem{allamanis2021self}
M.~Allamanis, H.~Jackson{-}Flux, and M.~Brockschmidt, ``Self-supervised bug detection and repair,'' in \emph{Advances in Neural Information Processing Systems 34: Annual Conference on Neural Information Processing Systems 2021, NeurIPS 2021, December 6-14, 2021, virtual}, 2021, pp. 27\,865--27\,876.

\bibitem{xu2023well}
W.~Xu, L.~Chen, C.~Su, Y.~Guo, Y.~Li, Y.~Zhou, and B.~Xu, ``How well static type checkers work with gradual typing? a case study on python,'' in \emph{2023 IEEE/ACM 31st International Conference on Program Comprehension (ICPC)}.\hskip 1em plus 0.5em minus 0.4em\relax IEEE, 2023, pp. 242--253.

\bibitem{goutte2005probabilistic}
C.~Goutte and {\'{E}}.~Gaussier, ``A probabilistic interpretation of precision, recall and \emph{F}-score, with implication for evaluation,'' in \emph{Advances in Information Retrieval, 27th European Conference on {IR} Research, {ECIR} 2005, Santiago de Compostela, Spain, March 21-23, 2005, Proceedings}, ser. Lecture Notes in Computer Science, vol. 3408.\hskip 1em plus 0.5em minus 0.4em\relax Springer, 2005, pp. 345--359.

\bibitem{li2020deep}
G.~Li, H.~Liu, J.~Jin, and Q.~Umer, ``Deep learning based identification of suspicious return statements,'' in \emph{27th {IEEE} International Conference on Software Analysis, Evolution and Reengineering, {SANER} 2020, London, ON, Canada, February 18-21, 2020}.\hskip 1em plus 0.5em minus 0.4em\relax {IEEE}, 2020, pp. 480--491.

\bibitem{alrashedy2023learning}
K.~Alrashedy, V.~J. Hellendoorn, and A.~Orso, ``Learning defect prediction from unrealistic data,'' \emph{CoRR}, vol. abs/2311.00931, 2023.

\bibitem{richter2022learning}
C.~Richter and H.~Wehrheim, ``Learning realistic mutations: Bug creation for neural bug detectors,'' in \emph{15th {IEEE} Conference on Software Testing, Verification and Validation, {ICST} 2022, Valencia, Spain, April 4-14, 2022}.\hskip 1em plus 0.5em minus 0.4em\relax {IEEE}, 2022, pp. 162--173.

\bibitem{patra2021semantic}
J.~Patra and M.~Pradel, ``Semantic bug seeding: A learning-based approach for creating realistic bugs,'' in \emph{{ESEC/FSE} '21: 29th {ACM} Joint European Software Engineering Conference and Symposium on the Foundations of Software Engineering, Athens, Greece, August 23-28, 2021}.\hskip 1em plus 0.5em minus 0.4em\relax {ACM}, 2021, pp. 906--918.

\bibitem{yasunaga2020graph}
M.~Yasunaga and P.~Liang, ``Graph-based, self-supervised program repair from diagnostic feedback,'' in \emph{Proceedings of the 37th International Conference on Machine Learning, {ICML} 2020, 13-18 July 2020, Virtual Event}, ser. Proceedings of Machine Learning Research, vol. 119.\hskip 1em plus 0.5em minus 0.4em\relax {PMLR}, 2020, pp. 10\,799--10\,808.

\bibitem{ye2022selfapr}
H.~Ye, M.~Martinez, X.~Luo, T.~Zhang, and M.~Monperrus, ``{SelfAPR}: Self-supervised program repair with test execution diagnostics,'' in \emph{37th {IEEE/ACM} International Conference on Automated Software Engineering, {ASE} 2022, Rochester, MI, USA, October 10-14, 2022}.\hskip 1em plus 0.5em minus 0.4em\relax {ACM}, 2022, pp. 92:1--92:13.

\bibitem{cassano2023type}
F.~Cassano, M.~Yee, N.~Shinn, A.~Guha, and S.~Holtzen, ``Type prediction with program decomposition and fill-in-the-type training,'' \emph{CoRR}, vol. abs/2305.17145, 2023.

\bibitem{yee2023machine}
M.~Yee and A.~Guha, ``Do machine learning models produce {TypeScript} types that type check?'' in \emph{37th European Conference on Object-Oriented Programming, {ECOOP} 2023, July 17-21, 2023, Seattle, Washington, United States}, ser. LIPIcs, vol. 263.\hskip 1em plus 0.5em minus 0.4em\relax Schloss Dagstuhl - Leibniz-Zentrum f{\"{u}}r Informatik, 2023, pp. 37:1--37:28.

\bibitem{pradel2020typewriter}
M.~Pradel, G.~Gousios, J.~Liu, and S.~Chandra, ``Typewriter: Neural type prediction with search-based validation,'' in \emph{{ESEC/FSE} '20: 28th {ACM} Joint European Software Engineering Conference and Symposium on the Foundations of Software Engineering, Virtual Event, USA, November 8-13, 2020}.\hskip 1em plus 0.5em minus 0.4em\relax {ACM}, 2020, pp. 209--220.

\bibitem{wei2022typet5}
J.~Wei, G.~Durrett, and I.~Dillig, ``{TypeT5}: Seq2seq type inference using static analysis,'' in \emph{The Eleventh International Conference on Learning Representations, {ICLR} 2023, Kigali, Rwanda, May 1-5, 2023}, 2023.

\bibitem{pandi2020opttyper}
I.~V. Pandi, E.~T. Barr, A.~D. Gordon, and C.~Sutton, ``{OptTyper}: Probabilistic type inference by optimising logical and natural constraints,'' \emph{CoRR}, vol. abs/2004.00348, 2020.

\end{thebibliography}

\end{document}